\documentstyle[prl,aps,epsfig]{revtex}   
\tightenlines
\begin{document}
\def\ba{\begin{eqnarray}}
\def\ea{\end{eqnarray}}
\def\be{\begin{equation}}
\def\ee{\end{equation}}
\def\({\left(}
\def\){\right)}
\def\[{\left[}
\def\]{\right]}
\def\lagrange {{\cal L}}
\def\del {\nabla}
\def\d {\partial}
\def\Tr{{\rm Tr}}
\def\half{{1\over 2}}
\def\fourth{{1\over 8}}
\def\bibi{\bibitem}
\def\S{{\cal S}}
\def\H{{\cal H}}
\def\K{{\cal K}}
\def\xx{\mbox{\boldmath $x$}}
\def\jmin{{${\cal J}^-$\,}}
\newcommand{\phpr} {\phi'}
\newcommand{\gam}{\gamma_{ij}}
\newcommand{\sqgam}{\sqrt{\gamma}}
\newcommand{\delk}{\Delta+3{\cal K}}
\newcommand{\dph}{\delta\phi}
\newcommand{\om} {\Omega}
\newcommand{\dom}{\delta^{(3)}\left(\Omega\right)}
\newcommand{\rar}{\rightarrow}
\newcommand{\Rar}{\Rightarrow}
\newcommand{\labeq}[1] {\label{eq:#1}}
\newcommand{\eqn}[1] {(\ref{eq:#1})}
\newcommand{\labfig}[1] {\label{fig:#1}}
\newcommand{\fig}[1] {\ref{fig:#1}}
\def\gsim{ \lower .75ex \hbox{$\sim$} \llap{\raise .27ex \hbox{$>$}} }
\def\lsim{ \lower .75ex \hbox{$\sim$} \llap{\raise .27ex \hbox{$<$}} }
\newcommand\bigdot[1] {\stackrel{\mbox{{\huge .}}}{#1}}
\newcommand\bigddot[1] {\stackrel{\mbox{{\huge ..}}}{#1}}

\twocolumn[\hsize\textwidth\columnwidth\hsize\csname@twocolumnfalse
\endcsname

\title{Steady-State Eternal Inflation}
\author{
Anthony Aguirre$^{1}$ and Steven Gratton$^{2}$}
\address{
$^{1}$School of Natural Sciences, Institute for Advanced Study
Princeton, New Jersey 08540, USA\\ {\rm email: aguirre@ias.edu}} 
\medskip
\address{
$^2$Joseph Henry Laboratories, Princeton University, Princeton, New Jersey
08544, USA\\ {\rm email: sgratton@princeton.edu}}
\date{\today}
\maketitle

\begin{abstract}

	Since the advent of inflation, several theorems have been
proven suggesting that although inflation can (and generically does)
continue eternally into the future, it cannot be extended eternally
into the past to create a ``steady-state'' model with no initial time.
Here we provide a construction that circumvents these theorems and
allows a self-consistent, geodesically complete, and physically
sensible steady-state eternally inflating universe, based on the flat
slicing of de Sitter space.  This construction could be used as the
background space-time for creation events that form big-bang-like
regions, and hence could form the basis for a cosmology that is
compatible with observations and yet which avoids an initial
singularity or beginning of time.

\end{abstract}
\vskip .2in
]  

\section{Introduction}

	Following the discovery of the cosmic expansion, cosmology
became dominated by two alternative paradigms.  The first, the Big
Bang (BB), is based on general relativity applied to a physical system
obeying the Cosmological Principle (CP) of spatial homogeneity and
isotropy.  In the BB, the observed universe evolved in a finite time from
a dense singular state before which classical space and time did not
exist.  The second, the Steady-State (SS), is based on the {\em Perfect}
Cosmological Principle (PCP) that the statistical properties of the
universe are independent also of time.  In the SS, the universe always
has and always will exist in a state statistically like its current
one, and time has no beginning~\cite{bondi,hoyle}.

	Observations of the thermal microwave background and evolution
in quasars and galaxies turned most astronomers away from the SS, and
its proponents were forced make their models only ``quasi-steady'',
with expansion and contraction cycles explaining the observed
evolution~\cite{hoylebook}.  But while the SS has approached the BB,
the BB has also approached the SS, in the form of ``eternal
inflation'': there is a broad consensus among its architects that
inflation---now considered by many to be an indispensable part of the
BB cosmology---never ends once it begins~\cite{etin,vil83}.  Rather,
inflation always continues somewhere and continually spawns new
thermalized regions, creating a mixture of inflating and non-inflating
areas that approaches some quasi-steady-state distribution eternally
into the future\cite{hn}.

	The SS cosmology is appealing because it avoids an initial
singularity, has no beginning of time, and does not require an initial
condition for the universe.  This led some to hope that inflation
could be extended eternally into the past to likewise avoid these
unpalatable necessities\cite{linde1}.  Attempts to do this failed,
however, and these failures have motivated the formulation of several
singularity theorems attempting to show that under very general
assumptions inflating spaces must contain singularities, so that
inflation can be at best ``semi-eternal'' into the
future\cite{ssrev,bgv}.  The most recent such theorem, for example,
attempts to show that an observer following ``almost any'' geodesic
will have finite past proper time if its ``locally measured Hubble
constant'' always exceeds some positive minimum value~\cite{bgv},
implying that inflating spaces are generically past geodesically
incomplete.

	This is an odd result as it applies to---and hence appears to
forbid---the seemingly physically reasonable classical SS cosmology
(which can itself be considered a form of eternal inflation).  In this
paper we attempt to resolve this incongruity by carefully examining
the implications of the singularity theorems and providing a
construction that allows for a physically reasonable geodesically
complete eternally inflating space-time in which physical observers can
have indefinitely long past proper time.  In Sec.~\ref{sec-ss} we
examine the classical SS cosmology in light of the singularity
theorems, and show how to construct a self-consistent and physically
reasonable model with its essential features.  Eternal inflation is
based on the same (de Sitter) space-time as the SS cosmology, and in
Sec.~\ref{sec-ei} we show how our construction might be used to
formulate a viable truly eternal model of inflation with big-bang-like
regions embedded in an eternally inflating background. We conclude in
Sec.~\ref{sec-conc}.

\section{The Steady-State Universe}
\label{sec-ss}

Let us now review the classical SS model.  The backbone of this theory
is the PCP, which holds that an observer at a randomly chosen position
in space and time measures physical properties of the universe that
are isotropic and that are statistically indistinguishable from any
other such observer.  This principle places strong conditions on a
cosmology that satisfies it.  The spatial part of the PCP implies that
space-time can be described by the Friedmann-Robertson-Walker (FRW) metric
with scale factor $a$.  The measurable Hubble parameter $H \equiv
(1/a)da/dt \equiv \dot{a}/a$ must be a constant in cosmic time $t$,
implying exponential expansion.  The universe must be spatially flat,
lest there be a time-varying ratio of the curvature scale to the
Hubble radius. The physical matter density must also be constant in
time.  Because of the Hubble expansion, this implies that particles
must be continually created so as to keep this density constant. This
requires that the number of particles created in a given four-volume
be proportional to the four-volume itself. (Mechanisms for doing this
are generally considered somewhat artificial, detracting from the
aesthetic appeal of the SS; but this is not important for the present
argument.)

	A subtle question that should be asked at this point is: why
have we stated that the universe is expanding (rather than
contracting) and that particles are being created (not destroyed)?
This is necessary if the arrow of time (AOT) is to point in the
direction of entropy creation: to maintain a SS, both microscopic and
coarse-grained entropy must, on average, be created as the universe
expands at a constant rate per unit physical volume. In other words,
if either microscopic or coarse-grained entropy were created as the
universe contracted, the entropy {\em density} would change in time,
violating the PCP.  The ``thermodynamic AOT'' defined by entropy
creation is in turn linked to the electromagnetic AOT. The latter
specifies, for example, that while moving charges can emit radiation
(creating an asymptotically spherical outgoing wavefront that is
either eventually absorbed or propagates to infinity), an incoming
spherical wavefront cannot ``miraculously'' assemble into a local
electromagnetic field that can move charges. Conditions sufficient to
ensure this behavior are that the ``retarded'' rather than
``advanced'' potentials are appropriate (which is tied to the
thermodynamic AOT) and that there be no radiation incoming from
infinity (the ``Sommerfield radiation
condition'')~\cite{layzer,wheelfeyn}.  This last condition is discussed
below; for now note only that in a SS model the thermodynamic and
(hence) electromagnetic AOTs are explicitly linked to the direction of
the expansion.

The metric for an exponentially expanding FRW universe can be written:
\ba ds^2=-dt^2 + e^{2Ht} \(dx^2+dy^2+dz^2\), \labeq{flatmetric} \ea
with $H$ constant chosen positive so that $a=e^{Ht}$ increases as $t$
increases as required by the preceding argument. This is a well
studied metric: that of the flat slicing of de Sitter space~\cite{he}.
4-dimensional de Sitter space-time can be thought of as a hyberboloid in
5-dimensional Minkowski space; see Fig.~1.  One may coordinatize this
hyperboloid in a variety of ways to yield slices of constant time that
are 

\begin{figure}
\epsfig{file=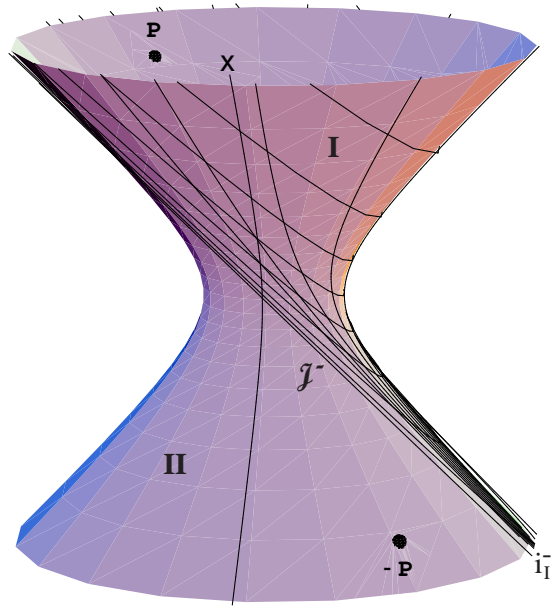,width=8cm}
\caption{A diagram of de Sitter space, with lines of equal time in the
flat slicing (the nearly diagonal parabolas) and comoving geodesics
(emerging from $i_{\rm I}^-$ in the lower-right) shown.  The
$t\rightarrow -\infty$ lightlike sufrace is labelled \jmin.  The
nearly vertical line labelled ``X'' represents a comoving geodesic in
the closed slicing, and ``P'' and ``-P'' are, respectively, points in
``region I'' (the portion of the space-time covered by the shown flat
slicing coordinates) and ``region II'' (the uncovered portion).
\label{fig-hyper}}
\end{figure}

\noindent open, flat or closed.  In our case the PCP has singled out
as physically appropriate the flat expanding system shown in Fig.~1:
slices of constant time are the nearly diagonal parabolas, with the
$t\rightarrow -\infty$ surface \jmin a null surface.  Comoving
observers emerge from past timelike infinity $i_{\rm I}^-$ in the
lower-right corner, with future light cones that open in the direction
of expansion, which is toward future timelike infinity at the top of
the diagram.

However, as we can see from the figure, this slicing does not cover
the entire hyperboloid.  So the universe described by~\eqn{flatmetric}
is geodesically incomplete; geodesics like that labelled ``X'' in
Fig.~1 can go ``through $t=-\infty$'' into the uncharted region. This
singularity at \jmin is exactly what the singularity
theorems~\cite{ssrev,bgv} point to, so let us look at the equations
here.  Consider a massive particle (the analogous argument can be made
for light).  Its geodesic equations in the metric~\eqn{flatmetric}
read \ba t''+ H e^{2Ht} {\mathbf{x}}'^2=0,~~~\(e^{2Ht}
{\mathbf{x}}'\)'=0, \ea where prime denotes a derivative with respect
to the proper time $\tau$ of the particle.  Integrating the latter
gives $e^{2Ht}\mathbf{x}'=\mathbf{v}$, with $\mathbf{v}$ constant.
Substituting into the former leads to \ba t'^2-e^{-2Ht}
{\mathbf{v}}^2=1, \labeq{timeges} \ea remembering the normalization of
the proper time.  Now consider the particular class of geodesics with
$\mathbf{v}=0$.  Then~\eqn{timeges} simply reads $t'^2=1$ so that
$\Delta t = \Delta \tau$ and for any value of $\tau$ the particle
stays in the region of space-time covered by the coordinates of the
flat slicing.  So these (comoving) geodesics are in fact complete in
the region of space-time described by~\eqn{flatmetric}.  Now consider
a geodesic with $\mathbf{v}\neq 0$.  Eq.~\eqn{timeges} leads to \ba
\int_{t_i}^{t_f} \frac{dt}{1+{\mathbf{v}}^2 e^{-2Ht}}=\tau_f-\tau_i.
\ea The LHS is unbounded as $t_f \rar +\infty$, showing that the
region of space-time described by Eq.~\eqn{flatmetric} is geodesically
complete to the future.  But it tends to a finite constant as $t_i
\rar -\infty$, implying that along this geodesic there is only a
finite proper time since $t=-\infty$.  Hence the region described by
Eq.~\eqn{flatmetric} is geodesically incomplete to the past.  This
sort of argument is the basis for the recent singularity theorem of
Ref.~\cite{bgv}.

What does this mean?  Is the steady-state model badly defined?  We
will address this question from several perspectives.  First, consider
what a particle $X$ on such a trajectory that came ``from the
outside'' into the region described by the metric~\eqn{flatmetric}
would look like to the geodesically complete comoving observer it
passes at time $t$.  It would appear to be a particle with energy $m_0
\sqrt{1+{\mathbf{v}}^2 e^{-2Ht}}$, where $m_0$ is its rest mass.  This
is time dependent and as $t \rightarrow -\infty$ particle $X$ has an
arbitrarily large energy.  That particles like $X$ are forbidden can
be seen in several complementary ways.  First, consider $X$ to be
propagating ``backwards in time'' toward \jmin.  If $X$ has {\em any}
nonzero interaction cross section with any particle in the universe
that has nonzero physical number density, then particle $X$ will
interact with an infinite number of them with arbitrarily high
energy. It would create, then, a ``spray'' of particles in a
light-cone opening toward \jmin, violating the CP to an abritrarily
great degree as \jmin is approached.\footnote{Note also that when
viewed forward in time, this requires collective, anti-entropic
behavior by an increasingly large number of particles as $t
\rightarrow -\infty.$} Now consider $X$ to be propagating forward in
time, starting ``from'' \jmin.  Then for the same reason, $X$ would
interact with an abitrarily large number of particles, yielding a
spray of high-energy particles filling a light-cone opening {\em away}
from \jmin, rendering any time-slice inhomogeneous.  Even supposing
the particle to somehow avoid all interactions, it would---simply by
virtue of its asymptotically infinite energy---still pick out a
preferred position, and violate the PCP.\footnote{A homogeneous family
of such incoming particles can satisfy the CP but only on one of the
flat spatial sections.}  In short, enforcing the PCP as $t \rightarrow
-\infty$ acts a boundary condition on \jmin that forbids any physical

\begin{figure}
\epsfig{file=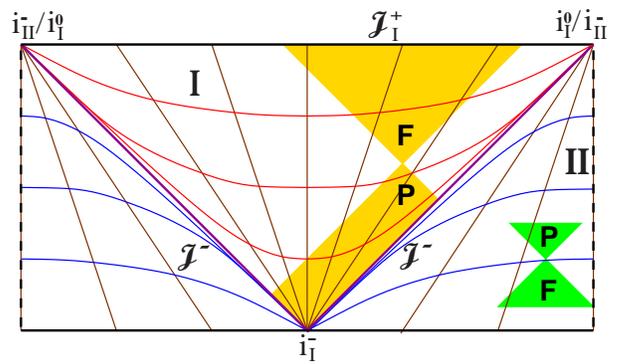,width=8cm}
\caption{Conformal diagram for de Sitter space.  Equal-time flat
slices are curved and spacelike; comoving geodesics are straight and
timelike.  The null surface \jmin represents $t\rightarrow -\infty$ in
both region I (above \jmin) and region II (below \jmin).  Shaded
regions represent future (``F'') and past (``P'') light cones, and
${\cal J}^+_{\rm I}$ is future timelike infinity for region I, while
$i_{\rm I}^-$ and $i_{\rm I}^0$ are its past-timelike and spacelike
infinities. The left and right (dotted) edges are identified.
\label{fig-conform}}
\end{figure}

\noindent particles from entering the SS universe from ``elsewhere''.
The only allowed physical things in the SS are
particles/photons/observers that are created within the space-time or
particles/observers that have world lines approaching the inextendible
comoving geodesics in their infinite past.

	Another way of looking at the situation is by asking what is
in the region on the other side of \jmin, labelled as ``region II'' in
Fig.~1.  Note first that no signal or particle created in region I can
escape into region II, because to do so it would have to travel along
a spacelike path, or somehow backward in time so as to exist before it
was created. Thus region II will see \jmin as a boundary from which no
particle or information emerges.  This is exhibited in the conformal
diagram for the flat slicing of de Sitter space-time (Fig.~2): the future
light-cone of any point in region I fails to intersect \jmin. Conceive
now some physical beings residing in region II.  In what sort of
universe do they live?  Consider first the electromagnetic AOT.  A
point in region II could {\em not} experience an incoming spherical
wave from infinity traveling along a light cone opening toward \jmin
(because no particles can emerge from there).  It could, however, send
such a wavefront {\em away} from \jmin.  This provides the Sommerfield
radiation condition in region II {\em as long as} the electromagnetic
AOT points away from \jmin, i.e. toward the {\em bottom} of Figs.~1
and~2, with comoving observers emerging from the point labelled
$i^-_{\rm II}$. Now, noting that the boundary condition on \jmin picks
out the flat slicing as preferred in region II, just as the PCP did in
region I, we are strongly motivated to apply the PCP in region II as
in region I (which would also link the thermodynamic and cosmological
AOTs as before).\footnote{It does not appear strictly necessary to
enforce the PCP in region II, though this makes the construction
simpler; all that is really necessary is that---as suggested but not
required by the Sommerfield condition---there is a globally
well-defined AOT that prevents particles created in region II from
passing through \jmin.  The PCP is a sufficient but probably not
necessary condition.}  Turning the page upside-down, we see that
region II now closely resembles region I, and that the ``no incoming
particles'' boundary condition on \jmin (which, recall, followed
simply from causality in region I) is exactly the necessary condition
to prevent the sort PCP-violating particles previously discussed in
the context of the geodesic completeness of region I.  And further,
the inability of particles created in region II to travel along
spacelike paths or into their own past prevents any particles from
traveling from region II into region I, completing the circle.

	In essence, this construction partitions the full de Sitter
space-time into a self-consistent set of two non-communicating SS
universes. An observer in region I does not see anything in its past
light cone from an observer in region II because that other observer
cannot signal into its past, and vice-versa.  Seen in this way the
boundary condition forbidding physical particles from following
geodesics across \jmin into one universe is in no way strange or
unreasonable, as it follows directly from the forbidding of causality
violations in the other universe. (One could similarly partition de
Sitter space-time by any non-timelike boundary ${\cal B}$ away from
which time flows. But \jmin is special: any spacelike ${\cal B}$ would
allow no eternal observers, and any other null ${\cal B}$ would be
less symmetric; moreover, \jmin is the only ${\cal B}$ that can be
``irrelevant'' by allowing interesting physics to occur throughout the
full de Sitter space-time even while no information flows from ${\cal
B}$.)

	The two universes resulting from the partition may not be
identical, despite sharing the null boundary, because $i_{\rm II}^-$
(which is the beginning point of all region II comoving geodesics) is
not necessarily the same as $i_{\rm I}^-$.  The universes can be {\em
made} identical, however, providing a more economical and perhaps more
elegant construction, through the identification of antipodal points
on the hyperboloid (demonstrated by equating the two points $P$ and
$-P$ in Fig.~1).  This identification maps \jmin onto itself.  The
resulting space-time (studied mathematically in Ref.~\cite{calabi})
contains no closed timelike curves, and in it \jmin is a surface of
infinitely early time that no physical particle can reach and from
which no physical particle can emerge.

Without the identification, the space-time manifold is time-orientable
in the mathematical sense that it is possible to continuously divide
non-spacelike vectors into two classes which can be labelled
``future'' and ``past''. In our construction these labels will only
correspond to the physical AOT in one of the two regions.  With the
identification the space-time is still a manifold but is not
mathematically time-orientable.  The physical AOT is, however, still
well-defined and no physical observer will see it reverse.

	While our construction is self-consistent for any physical
observer with an origin in the space-time, one might nevertheless ask
what a meta-physical invisible observer (with its own arrow of time)
might see as it follows for example one of the comoving geodesics of
the closed slicing which covers the full de Sitter space-time (shown
in Fig.~1 and labeled `X').  Moving in region I toward \jmin, it would
observe that the clocks on the comoving observers it passes (at huge
velocities as per Eq.~\eqn{timeges}) would read earlier and earlier
times, but that the clock of any one such observer would be turning at
an ever-slower rate.  Passing through \jmin it would see the {\em
times} diverge to minus infinity but the {\em rates} freeze.  Emerging
into region II (or in the identified case into another part of region
I), it would see the times increasing from minus infinity and the
rates increasing again.  It would perceive nothing singular happening,
interpreting the time reaching minus infinity and back as due to the
infinite length contraction between the comoving observers as its
speed relative to them momentarily becomes the speed of light.

\section{Eternal Inflation}
\label{sec-ei}

	The construction we have outlined gives either one or two
past- and future-eternally inflating regions of space-time, but is not
a viable cosmological model for the same (observational) reasons the
classical SS is not.  It could, however, provide the background for
events that create big-bang-like regions, one of which could describe
our environment.

	One possibility, based on ``old inflation''
~\cite{guth,guthwein}, was discussed by Vilenkin in Ref.~\cite{vil92}
but considered not to be viable because the background space-time (the flat
de Sitter slicing) is geodesically incomplete---the very problem to
which we have outlined a solution. To construct such a steady-state
inflation model, one may simply take the SS universe described above
and replace the particles by bubbles\footnote{Not all of the arguments
carry over directly.  For example, unlike the $X$ particle, a physical
observer (somehow) beginning in an inflating region could, without
encountering anything else, follow a timelike geodesic (necessarily
always passing through locally de Sitter space) toward \jmin and
(noticing nothing) pass through it.  The observer would then quickly
encounter a bubble and realize that it was traveling into the future
of its surrounding region.}
in which the scalar field will eventually roll down to the true
vacuum~\cite{bgt}; the interior of each bubble looks like an open FRW
cosmology to observers inside it\cite{coldel}. For a
suitably flat inflaton potential (as in ``open
inflation''\cite{bgt,openinf}), the FRW regions can be nearly flat,
homogeneous, and have scale-invariant density perturbations just as in
standard inflationary cosmology.  Like the particles in the SS
model, the number of bubble nucleation events in a given four-volume
is proportional to the four-volume
itself~\cite{coldel,col,calcol}. For non-overlapping bubbles, this
would yield a model obeying the PCP, as the physical number density of
the bubbles of a given size on each space-like surface would be
independent of time.  However, the bubbles do tend to overlap: given a
volume $V$ at a time $t$, bubbles formed after some earlier time $t_0$
at a rate $\lambda$ per unit 4-volume fill all of $V$ except a
fraction
\begin{equation}
f_{\rm inf} = \exp [-\lambda Q] = 
	\exp\left[{-4\pi\lambda(t-t_0) \over 3H^3}\right]
\labeq{finfl}
\end{equation}
for $(t-t_0) \gg H^{-1}$ and $V\rightarrow\infty$, where $Q$ is the
4-volume between $t_0$ and $t$ in the past light cone of a point in
$V$\cite{guthwein,vil92}.  Then $f_{\rm inf}\rightarrow 0$ as
$t_0\rightarrow -\infty$ and inflation seemingly halts.  However, as
argued in Ref.\cite{vil92} this is not necessarily the case.  One can
show that {\em if} a large but finite sphere of comoving volume
$\tilde V$ contains a fraction $f_{\rm inf} > 0$ of inflating volume,
then the inflating phase's physical volume increases with time in that
comoving volume, and its distribution at any time is a self-similar
fractal of dimension $D={3-{4\pi\lambda\over3H^4}}$ up to a scale $L
\propto \log (t-t_0)$ (that is, $V(r)f_{\rm inf}(r) \propto r^D$ for
$r < L$).  As $t_0 \rightarrow -\infty$ the distribution becomes
fractal on arbitrarily large scales and, because $D < 3$, the
inflating fraction of an arbitrarily large region tends to zero even
though parts inflate indefinitely. The global structure of the
space-time is still apparently de Sitter, however, as all inflating
regions are connected in space-time and the fractal ``skeleton'' of
inflating phase cannot be in any way affected by the regions of true
vacuum.  Note also that---although bubble intersections are
common---the interior of any given bubble formed at time $t$ will be
essentially homogeneous if $\lambda/H^4$ is small enough: it can be
shown~\cite{guthwein} that throughout all time, on average only
$(80\pi/9)(\lambda/H^4)$ bubbles formed prior to $t$ will intersect
the chosen bubble, and that the fractional volume of the chosen bubble
filled with bubbles formed after $t$ is also $(80\pi/9)(\lambda/H^4)$.

If the disquieting feature of this model that events of zero
probability occur in the universe (infinitely often) is
accepted,\footnote{Constructing a steady-state model is different from
constructing a model evolved from an initial condition in that one
must choose the state of the system at some particular time, and then
show that the state maintains itself and does not ``self-destruct''
by evolving into anything else.  Thus {\em assuming} that some region
is inflating is allowed (despite the odds) as long as this assumption
self-consistently leads to the same configuration at a later time.}
there is still cause for concern because an infinite fractal does {\em
not} satisfy the CP.  Note, however, that the distribution $f_{\rm
inf}(r)$ about each {\em inflating} point is the same, and is
independent of time.  Thus the inflating part of the universe does
satisfy a ``perfect'' (stationary in time) version of the
``Conditional Cosmographic Principle'' (proposed by
Mandelbrot\cite{mandel} as a generalization of the CP) that the
statistical distribution of inflating volume around any point is
isotropic and does not depend upon that point {\em on the condition}
that the point is itself inflating.  This principle, which holds for
both the inflating and thermalized regions, could therefore serve as a
replacement for the PCP in constructing an SS universe, though it is a
radical departure from the CP that implictly or explicitly lies at the
heart of almost all known cosmological models.

	The fractally inflating model just described is, however, not
the only conceivable possibility, and we can sketch out a number of
possible variations that could satisfy the usual CP and might (or
might not) improve upon it.

First, relaxing the PCP to the CP, the nucleation rate $\lambda$ could
tend to zero at ``early times'' so that the number of nucleation
events in the past light-cone of any point is finite.  This would
prevent $f_{\rm inf}$ from vanishing at any finite time (and the
fractal distribution would, as in future-eternal inflation, have an
outer scale above which it becomes homogeneous).  This would, however,
come at the great expense of introducing a preferred point in time.

Second, eternal inflation could occur in some 4+1 or
higher-dimensional manifold and somehow nucleate 3+1 dimensional
bubbles incapable of filling the space. This would also prevent
$f_{\rm inf}$ from vanishing.

Third and related, big-bang cosmologies could be formed within the
intersection of bubble walls in a higher-dimensional space-time.  A model
of this sort has been proposed in Ref.~\cite{buch}. As noted by
Ref.~\cite{bgv} this model as it stands is geodesically incomplete,
and so a construction such as that proposed here might be applied.

Fourth, thermalized or slowly-rolling regions could re-enter
inflation.  In the ``recycling'' scenario of Ref.~\cite{recycle} this
occurs due to quantum fluctuations of the inflaton.  There might be
some way in which the fraction of inflating space could be a finite
fraction of ``all'' space but is extremely difficult to see how to
compute this volume fraction in an unambiguous way.

Fifth, in the recent ``cyclic'' model of Ref.~\cite{cyclic} the
universe consists of two repeatedly colliding 3+1 dimensonal flat
branes embedded in a 4+1 dimensional bulk.  An (indestructible)
observer on one of the branes sees only a flat space which is almost
always exponentially expanding, and in which particles are
periodically created when an ``ekpyrotic'' collision occurs between
the branes. Averaged over sufficient time, the expansion is
exponential and the model comes to resemble a quasi-steady-state to an
observer on the brane.  The argument of Ref.~\cite{bgv} applies to
geodesics on the brane, all of which cannot then be fully extended
(because they encounter \jmin a finite proper time in their past.)
Here as in the classical SS all the matter that is created in this
model (in the brane collisions) originates at rest in the comoving
frame (defined here by those collisions), so no physical particles
follow the geodesics intersecting.  Nevertheless, as in the SS the
construction described herein may aid in constructing a global,
geodesically complete space-time for this scenario.

\section{Conclusion}
\label{sec-conc}

We have argued that the geodesically incomplete flat slicing of de
Sitter space-time can be completed in a self-consistent and physically
sensible way by considering it to be one of two similar or identical
regions of a full de Sitter space-time that is partitioned by the flat
slicing $t\rightarrow -\infty$ null surface \jmin.  Our construction
follows naturally from causality constraints which forbid each region
from sending particles or information into the other region.  It also
suggests intimate links between the arrows of time, the cosmic
expansion, and the (perfect) cosmological principle.

Although our arguments have been largely classical, they may have
interesting implications for the formulation of quantum field theories
(QFTs) in de Sitter space.  Both the ``two universe'' and identified
models are geodesically complete and seem therefore to provide a more
satisfactory background for QFTs than would an eternally inflating
spacetime with a boundary.  How (and whether) this quantum mechanical
formulation can be achieved constitutes an interesting subject for
future research.

Our construction may be applied to extend standard theories of
future-eternal inflation into the eternal past, though we do not claim
that such models are problem-free.  In particular, on any equal-time
slice the inflating regions form a fractal distribution of infinite
volume and yet a vanishing volume {\em fraction}, and the cosmological
principle must be replaced by some sort of ``conditional''
cosmological principle that can hold for infinite fractals.  For those
deeming these features undesirable, we have listed a number of
possible ways in which they might be avoided by other models.

Speaking more generally, what makes constructing eternal models of the
universe both appealing and difficult is that almost all, at bottom,
have the same essential features.  To be avoid a preferred time (as
seems highly desirable), the model must enforce some sort of {\nobreak
(quasi-)steady-state.} For the 2nd law of thermodynamics to hold
universally, the universe must then expand lest it be always in
equilibrium, and to be (quasi-)steady this expansion must be
(quasi-)exponential. Though not rigorous, these arguments lead
somewhat unavoidably to the flat slicing of de Sitter space-time or
some variant of it.\footnote{Chaotic ``eternal''
inflation~\cite{lindeei} may conceivably be an exception to this, as
the the inflaton potential in the inflating region tends to approach
the Planck energy and it is unclear how one is to talk about a global
structure of the universe at all.} Thus we suspect that a construction
like that proposed here may be necessary in any reasonable model for
an eternal universe that avoids a beginning of time.

\medskip
\centerline{\bf Acknowledgements}
\medskip

We thank Paul Steinhardt, Neil Turok and Alex Vilenkin for useful
discussions and comments.  AA is supported by a grant from the
W.M. Keck foundation.  SG is supported in part by US Department of
Energy grant DE-FG02-91ER40671.

\end{document}